\documentclass[12pt,a4paper,final]{revtex4}

\usepackage{lmodern}
\usepackage{sidecap}
\usepackage{ulem}
\usepackage{epsfig}
\usepackage{amsmath,amssymb,amsthm}
\usepackage{graphicx}
\usepackage{bm}
\usepackage{color}

\usepackage{enumerate}

\usepackage{float}
\usepackage{tabularx}

\DeclareMathAlphabet{\mathpzc}{OT1}{pzc}{m}{it}

\begin{document}

\renewcommand{\textfraction}{0.00}


\newcommand{\vAi}{{\cal A}_{i_1\cdots i_n}}
\newcommand{\vAim}{{\cal A}_{i_1\cdots i_{n-1}}}
\newcommand{\vAbi}{\bar{\cal A}^{i_1\cdots i_n}}
\newcommand{\vAbim}{\bar{\cal A}^{i_1\cdots i_{n-1}}}
\newcommand{\htS}{\hat{S}}
\newcommand{\htR}{\hat{R}}
\newcommand{\htB}{\hat{B}}
\newcommand{\htD}{\hat{D}}
\newcommand{\htV}{\hat{V}}
\newcommand{\cT}{{\cal T}}
\newcommand{\cM}{{\cal M}}
\newcommand{\cMs}{{\cal M}^*}
\newcommand{\vk}{\vec{\mathbf{k}}}
\newcommand{\bk}{\bm{k}}
\newcommand{\kt}{\bm{k}_\perp}
\newcommand{\kp}{k_\perp}
\newcommand{\km}{k_\mathrm{max}}
\newcommand{\vl}{\vec{\mathbf{l}}}
\newcommand{\bl}{\bm{l}}
\newcommand{\bK}{\bm{K}}
\newcommand{\bb}{\bm{b}}
\newcommand{\qm}{q_\mathrm{max}}
\newcommand{\vp}{\vec{\mathbf{p}}}
\newcommand{\bp}{\bm{p}}
\newcommand{\vq}{\vec{\mathbf{q}}}
\newcommand{\bq}{\bm{q}}
\newcommand{\qt}{\bm{q}_\perp}
\newcommand{\qp}{q_\perp}
\newcommand{\bQ}{\bm{Q}}
\newcommand{\vx}{\vec{\mathbf{x}}}
\newcommand{\bx}{\bm{x}}
\newcommand{\tr}{{{\rm Tr\,}}}
\newcommand{\pT}{p_\perp}
\newcommand{\RAA}{R_{AA}}
\newcommand{\sNN}{s_{\mathrm{NN}}}
\newcommand{\dif}{\mathrm{d}}
\newcommand{\lton}{\mathrel{\lower.9ex \hbox{$\stackrel{\displaystyle<}{\sim}$}}}
\newcommand{\bc}{\textcolor{blue}}
\newcommand{\red}[1]{\textcolor{red}{#1}}

\newcommand{\beq}{\begin{equation}}
\newcommand{\eeq}[1]{\label{#1} \end{equation}}
\newcommand{\ee}{\end{equation}}
\newcommand{\bea}{\begin{eqnarray}}
\newcommand{\eea}{\end{eqnarray}}
\newcommand{\beqar}{\begin{eqnarray}}
\newcommand{\eeqar}[1]{\label{#1}\end{eqnarray}}

\newcommand{\half}{{\textstyle\frac{1}{2}}}
\newcommand{\ben}{\begin{enumerate}}
\newcommand{\een}{\end{enumerate}}
\newcommand{\bit}{\begin{itemize}}
\newcommand{\eit}{\end{itemize}}
\newcommand{\ec}{\end{center}}
\newcommand{\bra}[1]{\langle {#1}|}
\newcommand{\ket}[1]{|{#1}\rangle}
\newcommand{\norm}[2]{\langle{#1}|{#2}\rangle}
\newcommand{\brac}[3]{\langle{#1}|{#2}|{#3}\rangle}
\newcommand{\hilb}{{\cal H}}
\newcommand{\pleft}{\stackrel{\leftarrow}{\partial}}
\newcommand{\pright}{\stackrel{\rightarrow}{\partial}}

\newcommand{\squeezeup}{\vspace{-2.5mm}}

\newcommand{\RomanNumeralCaps}[1]
    {\MakeUppercase{\romannumeral #1}}


\title{Jet-temperature anisotropy revealed through high-$\pT$ data}

\date{\today}

\author{Stefan Stojku}
\affiliation{Institute of Physics Belgrade, University of Belgrade, Serbia}

\author{Jussi Auvinen}
\affiliation{Institute of Physics Belgrade, University of Belgrade, Serbia}

\author{Lidija Zivkovic}
\affiliation{Institute of Physics Belgrade, University of Belgrade, Serbia}

\author{Pasi Huovinen}
\affiliation{Institute of Physics Belgrade, University of Belgrade, Serbia}
\affiliation{Incubator of Scientific Excellence---Centre for Simulations of
             Superdense Fluids, University of Wroc\l{}aw, Poland}

\author{Magdalena Djordjevic\footnote{E-mail: magda@ipb.ac.rs}}
\affiliation{Institute of Physics Belgrade, University of Belgrade, Serbia}

\begin{abstract}
  We explore to what extent, and how, high-$\pT$ data and predictions
  reflect the shape and anisotropy of the QCD medium formed in
  ultrarelativistic heavy-ion collisions. To this end, we use our
  recently developed DREENA-A framework, which can accommodate any
  temperature profile within the dynamical energy loss formalism. We
  show that the ratio of high-$\pT$ $v_2$ and $(1-R_{AA})$ predictions
  reaches a well-defined saturation value, which is directly
  proportional to the time-averaged anisotropy of the evolving QGP, as
  seen by the jets.
\end{abstract}

\maketitle

\section{Introduction}

Quark-gluon plasma is a new form of matter created in ultrarelativistic
heavy ion collisions. The main goal of relativistic heavy ion
physics~\cite{probe1,probe2,probe3,probe4} is to explore the
properties of this new form of matter~\cite{QGP1,QGP2}, which will in
turn lead to a fundamental understanding of QCD matter at its basic
level. In the QGP tomography approach that we advocate, the energy
loss of rare high energy partons traversing the medium is used to map
its properties.

We previously argued~\cite{shapeQGP} that, at large enough values of
transverse momentum (high-$\pT$), the ratio of the elliptic flow parameter $v_2$
and $1-\RAA$, where $\RAA$ is the nuclear suppression factor,
saturates, and reflects only the geometry of the system. This argument
was based on analytic considerations and a simple 1-dimensional
expansion~\cite{Bjorken,DREENAB}. To see how the evolving shape of the
collision system is reflected in the high-$\pT$ observables, we here study
the behavior of $v_2/(1-\RAA)$ in a system that expands both in
longitudinal and transverse directions.

\begin{figure}[h]
  \includegraphics[width=0.6\linewidth]{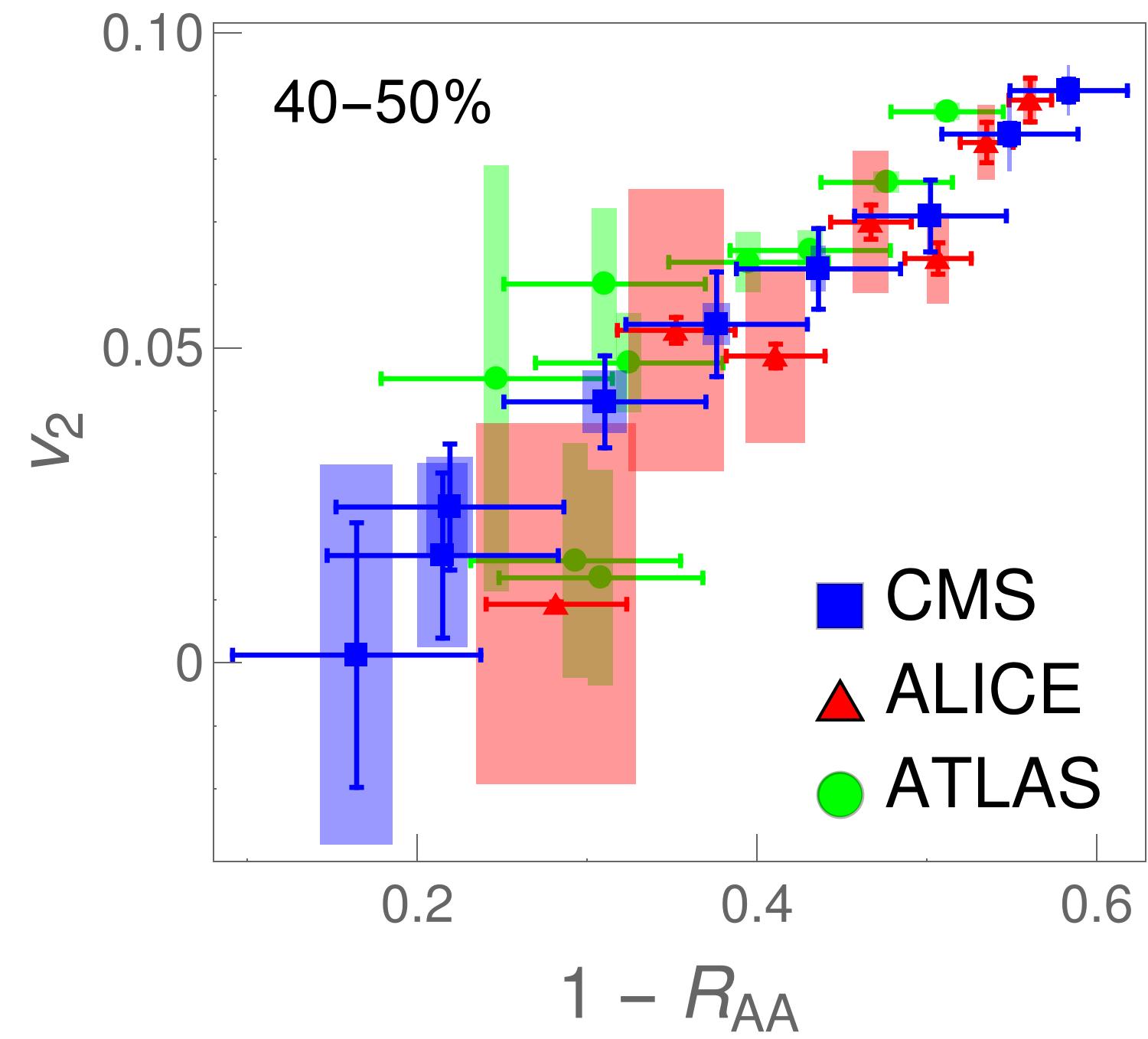}
  \caption{$v_2$ vs $1\!-\!\RAA$ for $\pT\!>\!10$~GeV data from 5.02 TeV Pb+Pb
    ALICE~\cite{ALICE_CH_RAA,ALICE_CH_v2} (red triangles),
    CMS~\cite{CMS_CH_RAA,CMS_CH_v2} (blue squares) and
    ATLAS~\cite{ATLAS_CH_RAA,ATLAS_CH_v2} (green circles) experiments. The
    data is shown for the 40-50\% centrality bin, while similar
    relation is obtained for other centralities. Each collaboration's
    datapoints correspond to different values of $\pT$, with both $v_2$
    and $1\!-\!\RAA$ decreasing with increasing $\pT$.}
  \label{Correlation}
\end{figure}

Furthermore, it has been experimentally observed that at high values
of transverse momentum $v_2$ and $1-\RAA$ are directly proportional. This is
shown in Fig.~\ref{Correlation} for ALICE~\cite{ALICE_CH_RAA,ALICE_CH_v2},
CMS~\cite{CMS_CH_RAA,CMS_CH_v2}, and ATLAS~\cite{ATLAS_CH_RAA,ATLAS_CH_v2}
data. Such relationship is equivalent to a $\pT$-independent ratio of
$v_2$ and $1-\RAA$, and we study whether the fluid dynamical calculation
can reproduce such proportionality, and whether we can relate this
observation to a physical property of the system, namely to
its anisotropy.

As is the standard approach in the study of ultrarelativistic
heavy-ion collisions, we assume the collision system to behave as a
locally thermalized dissipative fluid. It is well known that the
transverse expansion of such a system largely depends on the initial
gradients of the system, i.e., the initial state, and also on the
equation of state (EoS) and dissipative properties of the fluid. Thus,
to provide more general conclusions about the asymptotic behavior of
$v_2/(1-\RAA)$, it is necessary to explore not only one, but several,
different scenarios of fluid-dynamical evolution. On the other hand,
to provide physically meaningful results, tuning the
calculations to reproduce the observed low-$\pT$ data is necessary. As known,
finding suitable parameter combinations is a considerable task, and
therefore we do not explore the parameter space ourselves but employ
various initializations, and corresponding parameter sets, presented
in the literature. In particular, we restrict ourselves to models
tuned to reproduce the Pb+Pb collisions at $\sqrt{\sNN} = 5.02$ TeV
collision energy at the LHC. We further note that constraining the
calculation to reproduce the low-$\pT$ data does not guarantee the
reproduction of the high-$\pT$ data. Simultaneous reproduction of both
is not an aim of this study, but left for future work.

We will see that the temperature evolution in different evolution
scenarios is different enough to lead to observable differences in
high-$\pT$ $v_2$ and $1-\RAA$, and the ratio of these observables is
directly related to a suitably defined measure of the system anisotropy.

\section{Methods}

\subsection{Medium Evolution}

Our starting point and reference is a simple optical Glauber model
based initialization, which we use at different initial times $\tau_0
= 0.2,\, 0.4,\, 0.6,\, 0.8$ and $1.0$ fm. The initialization and code
used to solve viscous fluid-dynamical equations in 3+1 dimensions are
described in detail in Ref.~\cite{Molnar:2014zha}, and parameters to
describe Pb+Pb collisions at $\sqrt{\sNN} = 5.02$ TeV in
Ref.~\cite{Stojku:2020wkh}.  In particular, we use a constant shear
viscosity to entropy density ratio $\eta/s = 0.12$, and the EoS
parametrization $s95p$-PCE-v1~\cite{Huovinen:2009yb}.

Our second option, Glauber + Free streaming, is to use the Glauber model
to provide the initial distribution of (marker) particles, allow the
particles to stream freely from $\tau = 0.2$ to $1.0$ fm, evaluate the
energy-momentum tensor of these particles, and use it as the initial
state of the fluid. We evolve the fluid using the same code as in
the case of pure Glauber initialization. The EoS is $s95p$-PCE175,
i.e., a parametrization with $T_{\mathrm{chem}} = 175$
MeV~\cite{Niemi:2015qia}, and temperature-independent $\eta/s = 0.16$.
For further details, see Ref.~\cite{Stojku:2020wkh}.

As more sophisticated initializations, we employ EKRT, IP-Glasma
and T$_\mathrm{R}$ENTo. The EKRT model~\cite{Eskola:1999fc,
  Paatelainen:2012at,Paatelainen:2013eea} is based on the NLO
perturbative QCD computation of the transverse energy and a gluon
saturation conjecture. We employ the same setup as used in
Ref.~\cite{Auvinen:2020mpc} (see also~\cite{Niemi:2015qia}), compute an
ensemble of event-by-event fluctuating initial density distributions,
average them, and use this average as the initial state of the fluid
dynamical evolution. We again use the code of Molnar et
al.,~\cite{Molnar:2014zha}, but restricted to boost-invariant
expansion. The shear viscosity over entropy density ratio is
temperature dependent with favored parameter values from the Bayesian
analysis of Ref.~\cite{Auvinen:2020mpc}. Initial time is
$\tau_0 = 0.2$ fm, and the EoS is the $s83s_{18}$ parametrization
from Ref.~\cite{Auvinen:2020mpc}.

IP-Glasma model~\cite{ipglasma,ipglasma2} is based on Color Glass
Condensate~\cite{Iancu:2002xk,Iancu:2003xm,Gelis:2010nm,Lappi:2010ek}.
It calculates the initial state as a collision of two color glass
condensates and evolves the generated fluctuating gluon fields by
solving classical Yang-Mills equations. The calculated event-by-event
fluctuating initial states~\cite{Schenke:2020mbo} were further
evolved~\cite{Chun-private} using the MUSIC code~\cite{Schenke:2010nt,
  Schenke:2010rr,Schenke:2011bn} constrained to boost-invariant
expansion. We subsequently averaged the evaluated temperature profiles
to obtain one average profile per centrality class. In these
calculations, the switch from Yang-Mills to fluid-dynamical evolution
took place at $\tau_{\mathrm{switch}}=0.4$ fm, shear viscosity over
entropy density ratio was constant $\eta/s=0.12$, and the temperature-dependent bulk viscosity coefficient over entropy density ratio had
its maximum value $\zeta/s = 0.13$. The equation of state was based on
the HotQCD lattice results~\cite{HotQCD:2014kol} as presented in
Ref.~\cite{Moreland:2015dvc}.

T$_\mathrm{R}$ENTo~\cite{Moreland:2014oya} is a phenomenological model
capable of interpolating between wounded nucleon and binary collision
scaling, and with a proper parameter value, of mimicking the EKRT and
IP-Glasma initial states. As with the EKRT initialization, we create
an ensemble of event-by-event fluctuating initial states, sort them
into centrality classes, average, and evolve these average initial
states. Unlike in other cases, we employ the version of the VISH2+1
code~\cite{Song:2007ux} described in Refs.~\cite{Bernhard:2018hnz,
  Bernhard:2019bmu}. We run the code using the favored values of the
Bayesian analysis of Ref.~\cite{Bernhard:2019bmu}; in particular, allow
free streaming until $\tau = 1.16$ fm, the minimum value of the
temperature-dependent $\eta/s$ is 0.081, and the maximum value of the bulk
viscosity coefficient $\zeta/s$ is 0.052. The EoS is the same
HotQCD lattice results~\cite{HotQCD:2014kol} based parametrization as
used in Refs.~\cite{Bernhard:2018hnz, Bernhard:2019bmu}.

All these calculations were tuned to reproduce, in minimum, the
centrality dependence of the charged particle multiplicity, $\pT$
distributions and $v_2(\pT)$ in Pb+Pb collisions at
$\sqrt{\sNN} = 5.02$~TeV collisions.

\subsection{Energy loss: DREENA-A}

To calculate high-$\pT$ $R_{AA}$ and $v_2$, we used
our DREENA-A framework~\cite{DREENA-A}, where 'DREENA' is a shortcut
for "Dynamical Radiative ENergy Loss Approach". 'A' stands for
Adaptive, meaning that arbitrary temperature profile can be included
as input in the framework. The underlying dynamical energy loss
formalism~\cite{MD_Dyn,MD_Coll}, which is implemented in DREENA-A, has
several important properties necessary for reliable high-$\pT$
predictions~\cite{Blagojevic_JPG}:
{\ i)} Contrary to widely used static
approximation~\cite{BDMPS,ASW,GLV,HT}, the QGP is modeled as a
{\it finite} size and temperature medium, consisting of dynamical
(i.e.,\ moving) partons.
{\it ii)} Generalized Hard-Thermal-Loop approach~\cite{Kapusta} is
used, through which infrared divergences are naturally
regulated~\cite{MD_Dyn,MD_Coll,DG_TM}.
{\it iii)} The same theoretical framework is applied to both
radiative~\cite{MD_Dyn} and collisional~\cite{MD_Coll} energy loss
calculations.
{\it iv)} Calculations are generalized to include running
coupling~\cite{MD_PLB}, and non-perturbative effects related to
chromo-electric and chromo-magnetic
screening~\cite{MD_MagnMass,Peshier}. The framework does not have free
parameters in the energy loss, i.e., all the parameters are fixed to
standard literature values. Consequently, it can fully utilize
different temperature profiles as the only input in DREENA-A. $R_{AA}$
and $v_2$ predictions, generated under the same formalism and
parameter set (and calculated in a conventional way, see e.g.~\cite{DREENA-A}), can thus be systematically compared to experimental
data to map out the bulk properties of QGP.

\subsection{Experimental data}
We compared our predictions with data from the Pb+Pb collisions at
$\sqrt{\sNN} = 5.02$~TeV analysed by the LHC experiments
ALICE~\cite{ALICE_CH_RAA,ALICE_CH_v2},
CMS~\cite{CMS_CH_RAA,CMS_CH_v2}, and
ATLAS~\cite{ATLAS_CH_RAA,ATLAS_CH_v2}.
We used $v_2$ measurement obtained with the scalar product method. Since
we are interested in the high $ \pT $ region, we considered data with $
\pT>10 $~GeV.
The $ \pT $ bins are chosen as in the $v_2$ measurements, and the
$ R_{AA} $ distributions are interpolated to the chosen binning. Since
CMS experiment
used coarser granularity in centrality for $ R_{AA} $ measurements,
i.e. 10-30\% and 30-50\%,
we assigned the values obtained from 10-30\% (30-50\%) to both 10-20\%
and  20-30\% (30-40\% and 40-50\%). Finally, combined uncertainties on
the
$ v_2/(1-R_{AA}) $ are calculated assuming that $ v_2 $
and $R_{AA}$ are correlated.

\section{Results and discussion}

\begin{figure}[h]
  \includegraphics[width=\linewidth]{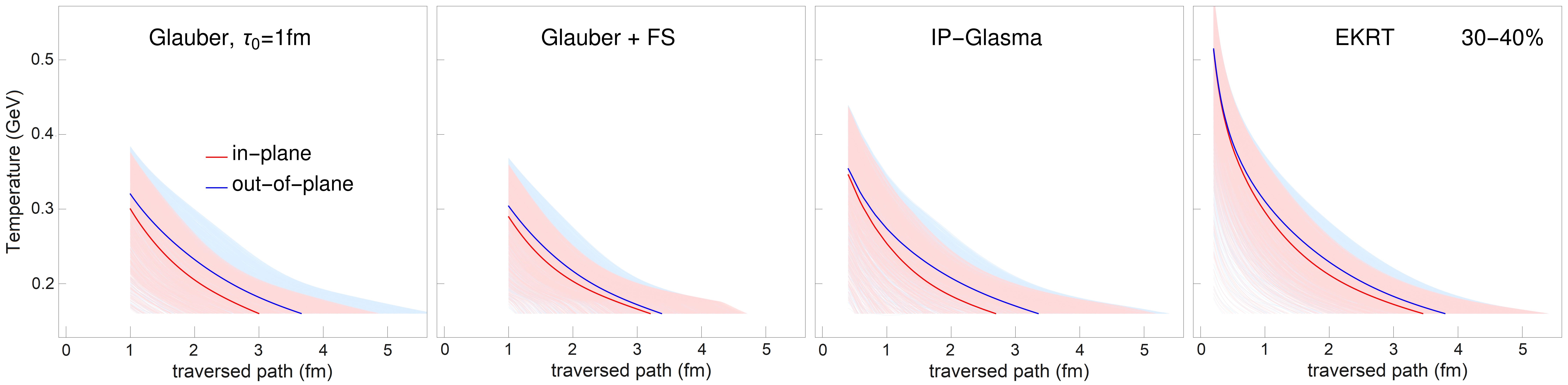}
  \caption{Light red (light blue) shaded areas represent the temperatures along the paths of high-$p_{\perp}$ partons when traversing the medium in the in-plane (out-of-plane) direction. For every scenario, we show 1250 in-plane and out-of-plane trajectories. The temperature profiles are from fluid-dynamical calculations of Pb+Pb collisions in the 30-40\% centrality class, utilising the Glauber model (with $\tau_0 = 1.0$ fm), Glauber + free streaming (FS), IP-Glasma, and EKRT initializations. Dark red (dark blue) curves represent the average temperature experienced by the particles in the in-plane (out-of-plane) directions.}
  \label{fig:Trajectories}
\end{figure}

To gain an intuitive insight into how different initializations
influence high-$\pT$ predictions, we show in Fig.~\ref{fig:Trajectories}
temperatures encountered by partons as a function of traversed
distance using four different temperature profiles. These plots are
produced by generating initial high-$p_{\perp}$ partons' positions
according to binary collision densities. Then these partons traverse
the medium in the in-plane (red) or out-of-plane (blue) directions,
and the temperature they experience is plotted as a function of their
path until they leave QGP. The larger the temperature that partons
experience while traversing the QGP, the larger the suppression in
high-$\pT$ observables. Similarly, a larger difference between
in-plane or out-of-plane temperatures is related to a larger
high-$\pT$ $v_2$.

From Fig.~\ref{fig:Trajectories}, we observe that partons traveling in
the in-plane and out-of-plane directions experience different
temperatures in different scenarios, and this leads to the different
behavior of high-$\pT$ particles. For example, based on the maximum
temperature encountered, we expect the largest suppression (i.e., the
smallest $R_{AA}$) for 'EKRT', while 'Glauber + FS' are expected to
lead to the largest $R_{AA}$. On the other hand, from the difference
in in-plane and out-of-plane temperatures, we expect the largest $v_2$
for 'Glauber, $\tau_0 = 1$~fm', followed by 'IP-Glasma', while $v_2$
for 'EKRT' should be notably smaller. The ordering of $\RAA$ and $v_2$
is thus different for different evolution scenarios, and therefore it is a priori unclear what the
ordering of $v_2/(1-R_{AA})$ might be. In this section, we aim to
address the following questions: {\it i}) Is the saturation in
$v_2/(1-R_{AA})$ at high-$\pT$ still observed for these different
profiles, as expected from our previous analytical arguments and simple
1D Bjorken expansion~\cite{shapeQGP}? {\it ii})
If yes, does this saturation carry information about the anisotropy of
the system, and {\it iii}) What kind of anisotropy measure corresponds
to the high-$\pT$ data?

\begin{figure}[h]
  \includegraphics[width=1.0\linewidth]{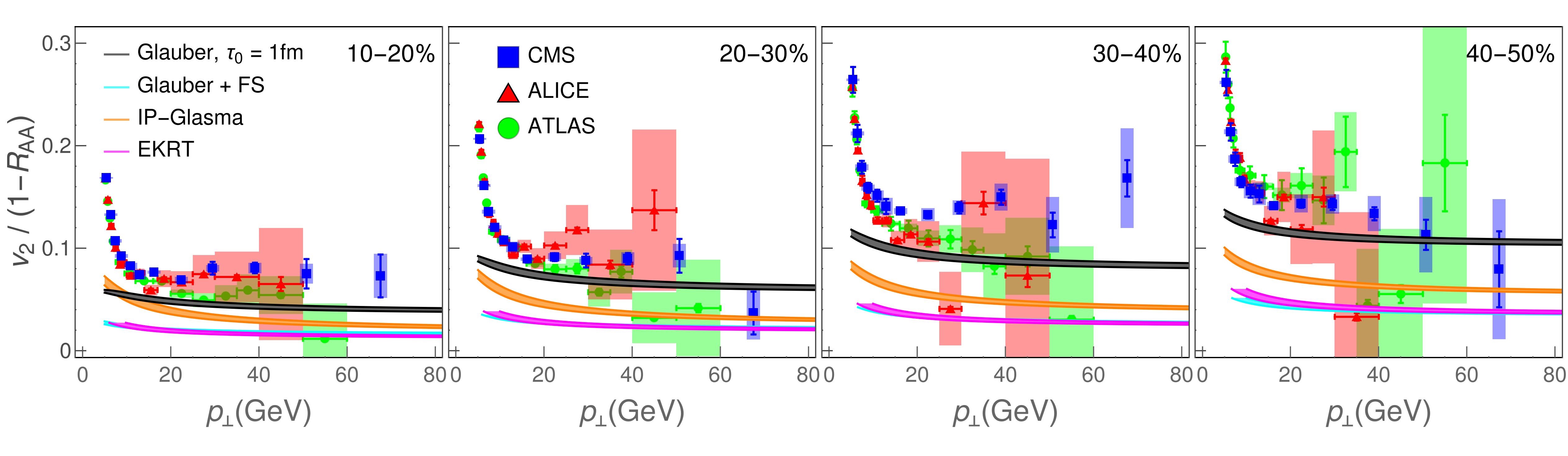}
  \caption{Theoretical calculations of the $v_2/(1-R_{AA})$ ratio as a
    function of transverse momentum $p_{\perp}$, calculated within
    DREENA-A framework using four different temperature profiles
    (Glauber with $\tau_0 = 1.0$ fm, Glauber + free streaming (FS),
    IP-Glasma, and EKRT). Theoretical predictions are compared with
    5.02 TeV Pb+Pb ALICE~\cite{ALICE_CH_RAA,ALICE_CH_v2} (red triangles),
    CMS~\cite{CMS_CH_RAA,CMS_CH_v2} (blue squares) and
    ATLAS~\cite{ATLAS_CH_RAA,ATLAS_CH_v2} (green circles) data. Each
    panel corresponds to a different centrality (10-20\%, 20-30\%,
    30-40\%, 40-50\%). The bands correspond to the uncertainty in the
    magnetic to electric mass ratio. The upper (lower) boundary of
    each band corresponds to $\mu_M / \mu_E = 0.4$~$(0.6)$.}
  \label{fig:panel}
\end{figure}

To start addressing these questions, we show in Fig.~\ref{fig:panel}
how $v_2/(1-\RAA)$ depends on $\pT$ for different temperature
profiles.~\footnote{To avoid cluttering the figure, we concentrate on the same
four profiles as shown in Fig.~\ref{fig:Trajectories}.} As seen, the
ratio is almost independent of $\pT$ above $\pT \approx 30$ GeV,
although IP-Glasma shows some $\pT$ dependence even above this
limit. We also confirmed that the saturation is obtained for other
hydrodynamic calculations outlined in Subsection IIA (not shown). Thus, the
phenomenon of $v_2/(1-\RAA)$ saturation is indeed robust, i.e., holds
for a variety of different transversally expanding systems.

We also observe that some profiles lead to better agreement with the
data compared to the others. However, the goal of this paper is not to
achieve the best agreement with the data (which is, in itself, a
considerable task). Instead, we want to explore how the differences in
the medium evolution are reflected through high-$\pT$ data, and leave
the comparison between predictions and the data to later studies (some
aspects were addressed in~\cite{Stojku:2020wkh}).

\begin{figure}[h]
  \includegraphics[width=0.67\linewidth]{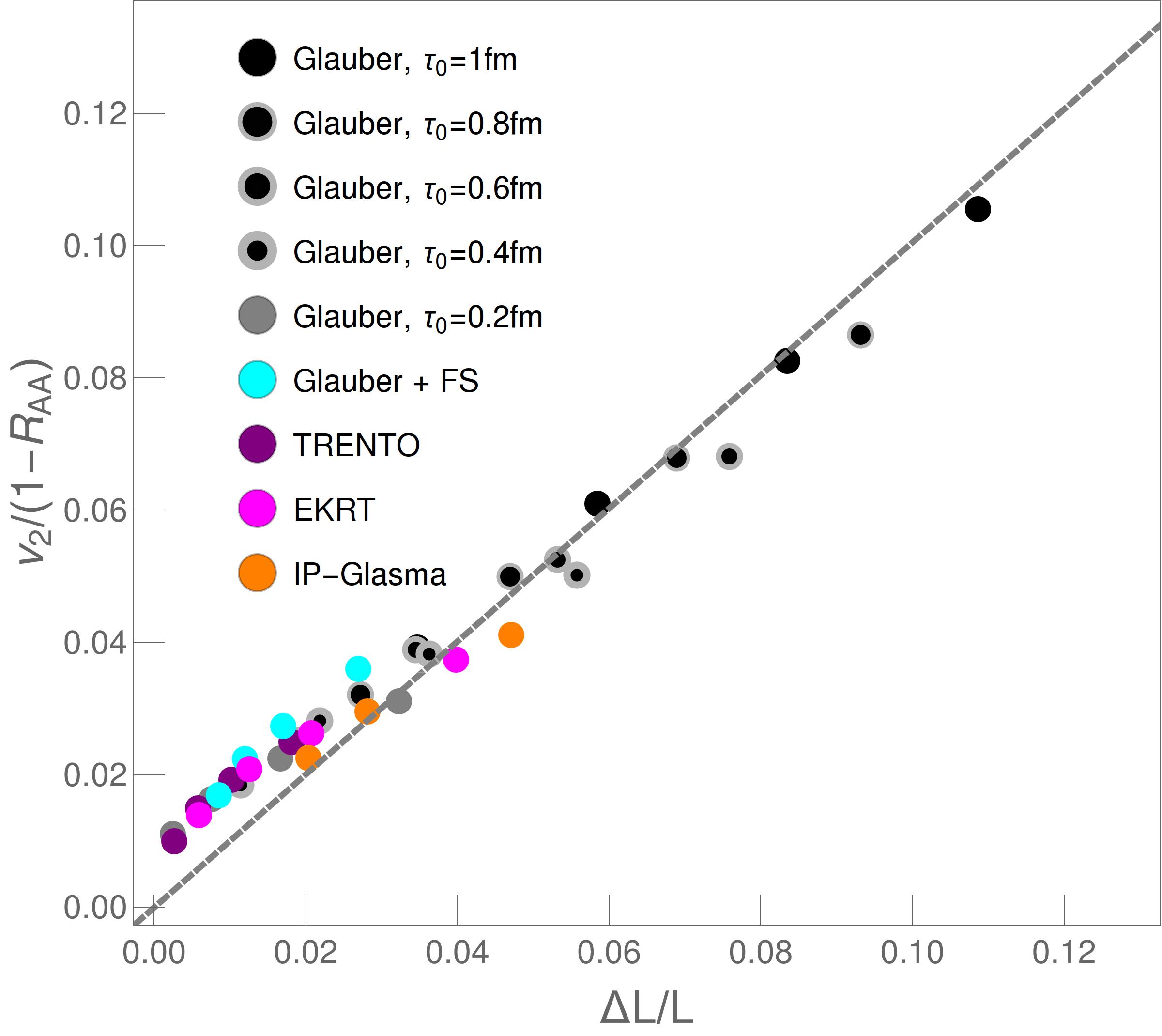}
  \caption{Charged hadron $v_2/(1-R_{AA})$ as a function of
    path-length anisotropies $\Delta L/L$, for various centrality
    classes and temperature profiles. The value of transverse momentum
    is fixed at $p_{\perp} = 100$ GeV. The linear fit yields a slope
    of approximately 1.}
  \label{fig:AnisotropyPlot}
\end{figure}

To find out whether the saturation values of $v_2/(1-\RAA)$ are
correlated with the system geometry, we evaluate the average path
length of partons, $\langle L\rangle$, and its anisotropy
\begin{align}
  \frac{\Delta L}{\langle L \rangle}
              = \frac{ \langle L_{out} \rangle - \langle L_{in} \rangle }
                     {\langle L_{out} \rangle + \langle L_{in} \rangle},
    \label{varSigmaDef}
\end{align}
where $\langle L_{in} \rangle$ and $\langle L_{out} \rangle$ refer to
the average path-length of high-$p_{\perp}$ particles in the in-plane
and out-of-plane directions. For every temperature profile, $\langle
L_{in} \rangle$ and $\langle L_{out} \rangle$ are calculated using the
Monte Carlo method to generate an initial hard parton position in the
XY plane according to the binary collision densities. The parton then
traverses the medium in the $\phi = 0$ (or $\phi = \pi/2$) direction,
until the temperature at parton's current position drops below critical
temperature $T_c$. We use $T_c$=160 MeV, which is within the uncertainty of
the lattice QCD critical temperature of $154 \pm 9$~MeV~\cite{Tcritical}. We then obtain
$\langle L_{in} \rangle$ and $\langle L_{out} \rangle$ by averaging
the in-plane and out-of-plane path lengths over many different partons.

In Fig.~\ref{fig:AnisotropyPlot} we have plotted the values of
$v_2/(1-R_{AA})$ evaluated at 100 GeV for different evolution scenarios vs.\ the corresponding
path-length anisotropies $\Delta L/\langle L \rangle$. Except for
IP-Glasma, each scenario is presented with four points corresponding
to the centrality classes 10-20\%, 20-30\%, 30-40\%, 40-50\%. We have
omitted the 40-50\% class of IP-Glasma since the average profile for
this centrality class was not smooth enough to produce reliable $v_2$
and $R_{AA}$ results.

Figure~\ref{fig:AnisotropyPlot} shows a surprisingly simple relation
between $v_2/(1-\RAA)$ and $\Delta L/\langle L \rangle$, where the
dependence is linear with a slope of almost 1. The saturation value is
therefore dominated by the geometry of the system, although at small
values of $\Delta L/\langle L \rangle$ there is a deviation from the
linear proportionality. It is worth noticing that here the values of
$\Delta L/\langle L \rangle$ are much smaller than in our earlier 1D
study~\cite{shapeQGP}. Also, even if the values of $\Delta L/\langle L
\rangle$ are very different for different initializations, the initial
anisotropies, $\epsilon_{2,2}$, are not so different. The general
trend is that the earlier the transverse expansion begins (fluid
dynamical or otherwise), the smaller the $\Delta L/\langle L \rangle$
in the same centrality class. The time it
takes the parton to reach the edge of the system is almost independent
of $\tau_0$, but small $\tau_0$ means that by the time the parton
reaches the edge, the system has evolved longer, and the initial
anisotropy has been diluted more. Thus, the earlier the expansion
begins, the lower the $\Delta L$, and $\Delta L/\langle L\rangle$
depicts the mentioned sensitivity on the time when expansion begins.

\begin{figure}[h]
  \includegraphics[width=0.67\linewidth]{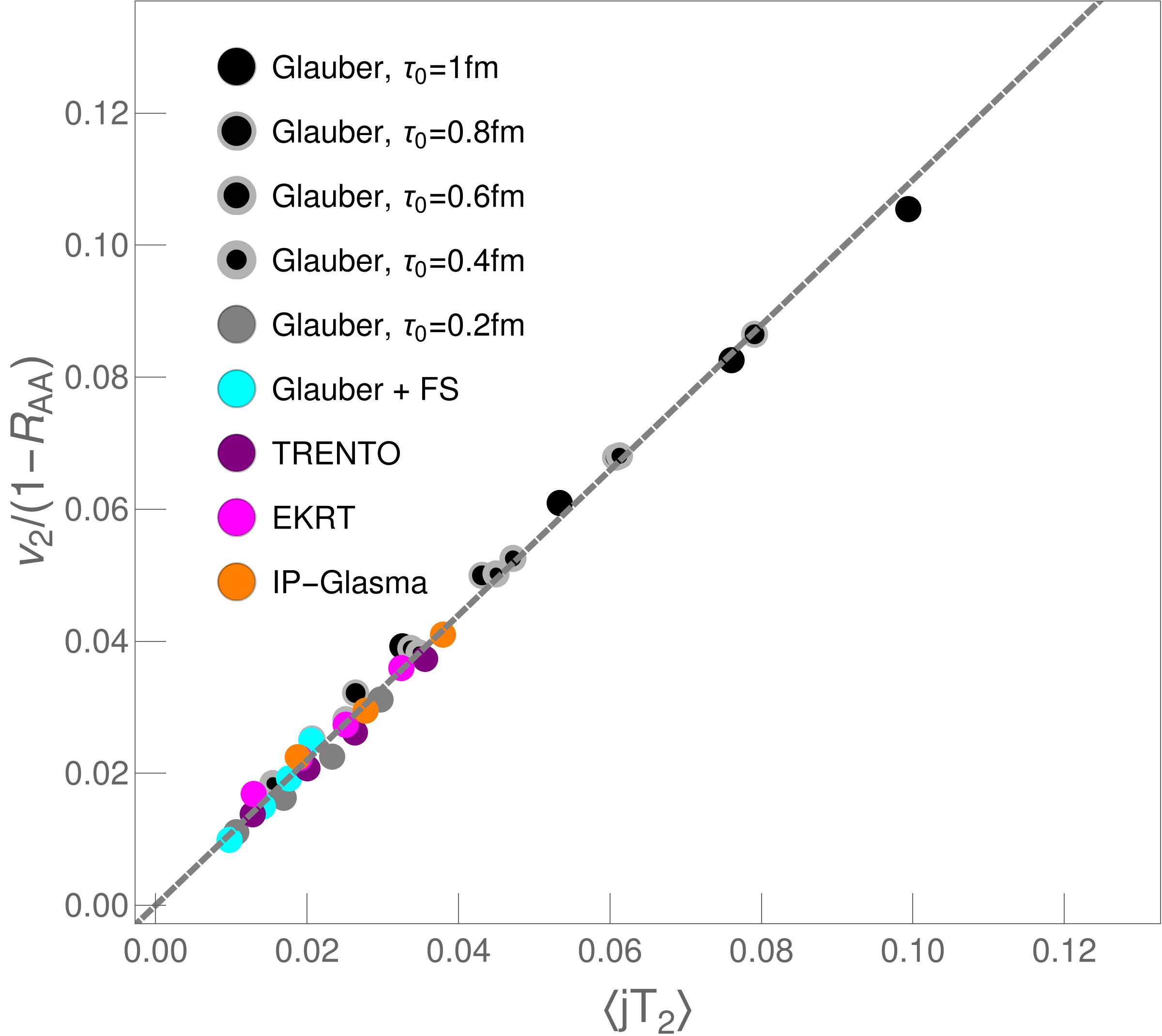}
  \caption{The values of $v_2/(1-R_{AA})$ of charged hadrons as a
    function of the average jet-temperature anisotropy $\langle
    jT_2\rangle$ for various centrality classes and temperature
    profiles. The value of transverse momentum is fixed at $p_{\perp} = 100$ GeV.
    The linear fit yields a slope of approximately 1.}
  \label{fig:PasiAnisotropyPlot}
\end{figure}

$\Delta L/\langle L \rangle$ depends on the shape of the system, and
how the shape evolves, but it is difficult to see how the evolution
affects it. Thus, we wanted to see if it is possible to define a more
direct measure of anisotropy, with an explicit dependence on
time-evolution. After testing various generalizations of the
conventional measure of the spatial anisotropy, $\epsilon_{m,n}$, we
ended up evaluating the average of temperature cubed encountered by
partons propagating with angle $\phi$ with respect to the reaction
plane:
\beq
  jT(\tau,\phi)
      \equiv \frac{\int\dif x \dif y\,T^3(x+\tau\cos\phi,y+\tau\sin\phi,\tau)\,n_0(x,y)}
                  {\int\dif x \dif y\,n_0(x,y)},
\ee
where $n_0(x,y)$ is the density of the jets produced in the primary
collisions, i.e., the density of the binary collisions. This
distribution is not azimuthally symmetric, and we may evaluate its second
Fourier coefficient:
\beq
  jT_2(\tau)
    = \frac{\int\dif x \dif y\,n_0(x,y)
             \int\! \dif\phi\cos 2\phi\,T^3(x+\tau\cos\phi,y+\tau\sin\phi,\tau)}
           {\int\dif x \dif y\,n_0(x,y)
             \int\!\dif\phi\,T^3(x+\tau\cos\phi,y+\tau\sin\phi,\tau)}.
\ee
Moreover, a simple time-average of $jT_2$,
\beq
\langle jT_2 \rangle
   = \frac{\int_{\tau_0}^{\tau_{\mathrm{cut}}} \dif\tau\,jT_2(\tau)}{\tau_{\mathrm{cut}}-\tau_0},
\label{jT2_aver}
\ee
where $\tau_{\mathrm{cut}}$ is defined as the time when the center of
the fireball has cooled down to critical temperature $T_c$, is
directly proportional to the ratio $v_2/(1-\RAA)$ as shown in
Fig.~\ref{fig:PasiAnisotropyPlot}.

We call this measure the average jet-temperature anisotropy of the
system. When we evaluate the ratio of $v_2/(1-\RAA)$ and $\langle
jT_2\rangle$ in the $\pT$ range where the $v_2/(1-\RAA)$ ratio has
saturated for all models ($\pT>80$~GeV), and average over all the cases shown in
Fig.~\ref{fig:PasiAnisotropyPlot} we obtain
\begin{equation}
  \frac{v_2/(1-\RAA)}{\langle jT_2\rangle} = 1.1 \pm 0.1.
 \label{eq:jT2_from_v2RAA}
\end{equation}
Thus, $v_2/(1-\RAA)$ at high-$\pT$ carries direct information of the
system geometry, and its anisotropy. Since unity is within one
standard deviation from the average, for practical purposes we use
approximation $v_2/(1-\RAA) = \langle jT_2\rangle$.

The above analysis was performed on charged hadrons. However, if $v_2/(1-\RAA)$ indeed reflects the anisotropy $\langle jT_2\rangle$, then the Eq.~(\ref{eq:jT2_from_v2RAA}) should be independent of flavor. Due to large mass, $R_{AA}$ and $v_2$ of heavy flavor
particles depend on $\pT$ differently from charged hadrons' $R_{AA}$
and $v_2$. To test whether the $v_2/(1-\RAA)$ ratio of heavy flavor
particles also saturates at high $\pT$, and whether
Eq.~(\ref{eq:jT2_from_v2RAA}) is valid for them, we performed the same
analysis on $R_{AA}$ and $v_2$ of D and B mesons.
We obtained that $v_2/(1-\RAA)$ indeed saturates at
$\pT>20$~GeV for B mesons and at $\pT>80$~GeV for D mesons, and that, after
saturation, the Eq.~(\ref{eq:jT2_from_v2RAA}) is robust for all types
of flavor (results not shown). This further supports that $v_2/(1-\RAA)$ at high-$\pT$
directly carries information on the medium property, as revealed
through this extensive analysis.

\begin{figure}[h]
  \includegraphics[width=0.75\linewidth]{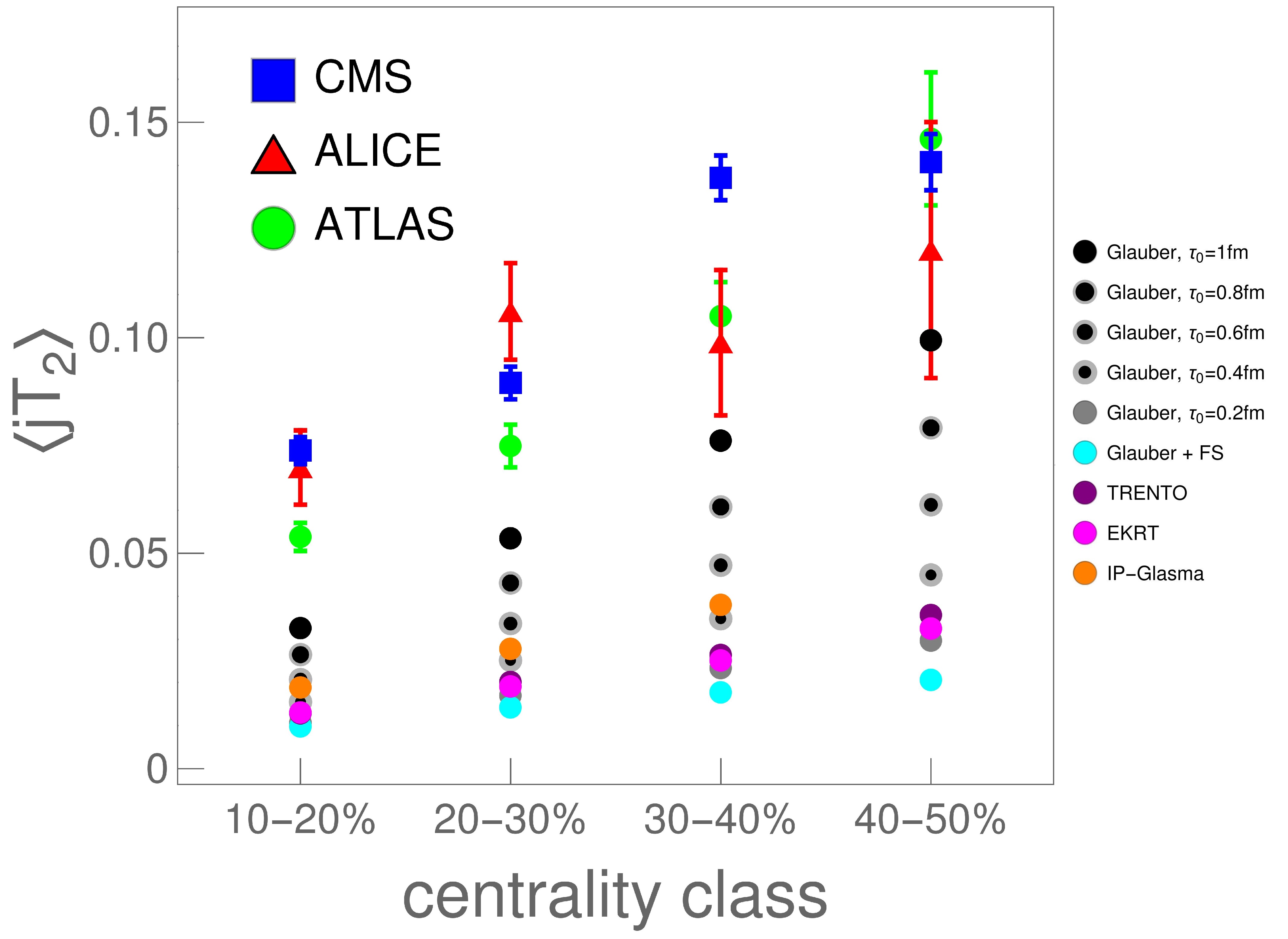}
  \caption{Constraints to jet-temperature anisotropy ($\langle jT_2\rangle$, shown on
    y-axis) evaluated from high-$p_{\perp}>$~20GeV $R_{AA}$ and
    $v_2$ experimental data using $\langle jT_2\rangle = v_2/(1-\RAA)$
    (see Eq.~\ref{eq:jT2_from_v2RAA}), for four
    different centrality regions (shown on x-axis): 10-20\%, 20-30\%,
    30-40\%, 40-50\%. 5.02 TeV Pb+Pb ALICE~\cite{ALICE_CH_RAA,ALICE_CH_v2}
    (red triangles), CMS~\cite{CMS_CH_RAA,CMS_CH_v2} (blue squares)
    and ATLAS~\cite{ATLAS_CH_RAA,ATLAS_CH_v2} (green circles) data are
    used. For each centrality, the experimental constraints are compared
    with the average jet-temperature anisotropy $\langle jT_2\rangle$
    for various evolution scenarios, indicated on the legend.}
  \label{jt2Plot}
\end{figure}

Finally, we evaluated the favored $\langle jT_2\rangle$ range from the experimentally measured $\RAA (\pT)$ and $v_2 (\pT)$ for charged hadrons at different
centralities. The $\langle jT_2\rangle$ values are obtained by fitting $v_2(\pT)/(1-\RAA(\pT))$ experimental data shown in Fig.~\ref{fig:panel} with a constant function for $ \pT>20 $~GeV using  MINUIT~\cite{MINUIT} package within ROOT~\cite{ROOT} code, taking uncertainties into account.
The fitted ratio was then converted to $\langle jT_2\rangle$ by assuming their equality.
As shown in Fig.~\ref{jt2Plot}, all three experiments
lead to similar values of $\langle jT_2\rangle$, though the uncertainty is still large.

We also note that $\langle jT_2\rangle$ is a bulk-medium property,
which can be directly evaluated from bulk-medium simulations through
Eqs.~(2)--(4), independently of high-$\pT$ data. Thus, experimental
data can be used to restrict the value of this quantity. In
Fig.~\ref{jt2Plot}, we see that none of the evolution scenarios tested
in this manuscript is in good agreement with the experimental data
(despite the above mentioned large uncertainty), i.e., lead to smaller
jet-temperature anisotropy than experimentally favored. We thus
show that jet-temperature anisotropy provides an important constraint
on bulk-medium simulations, and that future bulk-medium calculations
should be tuned to reproduce the experimentally constrained $\langle
jT_2\rangle$ as well. Moreover, in the high-luminosity $3^{rd}$ run at
the LHC, the error bars for $\RAA (\pT)$ and $v_2 (\pT)$ are expected
to be significantly reduced, which will subsequently lead to a notably
better experimental constraint of $\langle jT_2\rangle$, also enabling
better constraint on bulk-medium simulations.

\section{Summary}

In this study, we used our recently developed DREENA-A framework to
explore how the temperature evolution of the QGP droplet influences
high-$\pT$ $v_2/(1-R_{AA})$ predictions. The framework does not use
any free parameter within the energy loss model and consequently
allows to fully explore these profiles as the only input in the
model. We showed that saturation in $v_2/(1-R_{AA})$, clearly seen in the experimental data, is robustly
obtained for the comprehensive set of fluid-dynamical calculations
covered in this study, as well as different types of flavor. We further revealed that this saturation value
corresponds to a property of the system we defined as
average jet-temperature anisotropy $\langle jT_2\rangle$. We also
showed how to relate $\langle jT_2\rangle$ to experimental data,
providing a new important constraint on bulk-medium simulations.
None of the evolution scenarios that we tested here was in good agreement with experimentally inferred $\langle jT_2\rangle$ values, which argues that it is important to accordingly tune the bulk-medium simulations, particularly with the high-luminosity $3^{rd}$ run at the LHC. Our
approach demonstrates the utility of the QGP tomography, i.e., the
potential for extracting the bulk QGP properties jointly from low and
high-$\pT$ data.

{\em Acknowledgments:}
We thank Chun Shen and Harri Niemi for sharing their results with us.
This work is supported by the European Research Council, grant
ERC-2016-COG: 725741, and by the Ministry of Science and Technological
Development of the Republic of Serbia. PH was also supported by the program Excellence Initiative
Research University of the University of Wroc\l{}aw of the
Ministry of Education and Science.

\end{document}